# Sustainable wafer-scale integration of epitaxial ZnO on silicon for piezoelectric devices

D. Sanchez-Fuentes[1], R. Desgarceaux[1], A. Rahal[1], L. García[1], S. Bousri[1], S. Ding[1], N. Camara[1,2], F. Pascal[1], R. García-Bermejo[1], N. Guillaume[3], G. Ardila [3], J. Gazquez[4], C. Magen,[5,6] S Plana-Ruiz[7], C. Guasch[1], and A. Carretero-Genevrier[1]*

1. Institut d'Électronique et des Systèmes (IES), CNRS UMR 5214 – Université de Montpellier, Montpellier, France.).
2. EPF Graduate School of Engineering, Montpellier, France
3. Université Grenoble Alpes, Univ. Savoie Mont Blanc, CNRS, Grenoble INP, CROMA, F-38000 Grenoble, France
4. ICMAB, CSIC Campus UAB 08193 Bellaterra, Catalonia, Spain.
5. Instituto de Ciencia de Materiales de Aragón (ICMA), Universidad de Zaragoza-CSIC, Facultad de Ciencias, Universidad de Zaragoza, Pedro Cerbuna 12, 50009 Zaragoza, Spain
6. Laboratorio de Microscopías Avanzadas (LMA), Instituto de Nanociencia de Aragón (INA), Universidad de Zaragoza, Mariano Esquillor, Edificio I+D, 50018 Zaragoza, Spain
7. Servei de Recursos Científics i Tècnics, Universitat Rovira i Virgili, 43007 Tarragona, Catalonia, Spain.

E-mail: adrien.carretero@cnrs.fr

**Abstract**

To sustainably support the ongoing energetic transition, we need functional metal oxides capable of converting energy, and produce storage, and sensing devices. However, these materials suffer from a high economic cost of manufacturing, and their production in a sustainable way is, to date, a milestone. Additionally, the technical challenges, such as scalability and integration of silicon for industrial processing using microelectronic technologies, impose strict conditions for the entire materials process. In this work, we engineer α-quartz virtual substrates up to 4 inches facilitating the large-scale and sustainable integration of multifunctional epitaxial ZnO metal oxide microwire films on silicon. These materials are exclusively manufactured on silicon using solution chemistry, providing single-chip solutions that can meet strict economic constraints for developing sustainable devices at a lower cost. Through this integrative technology, we demonstrate the microfabrication of epitaxial (110)ZnO/(100)α-quartz/(100)silicon piezoelectric membrane resonators at the wafer-scale with potential applications in energy conversion and sensing. We combined four dimensional (4D)-STEM diffraction technology and Piezoelectric Force Microscopy (PFM) to establish a correlation between out of plane crystalline strain and piezoelectric response in epitaxial (110) ZnO at the microscale. Finally, we proved the fabrication of 800 nm thick (110) ZnO suspended membranes that can be transferred to flexible substrates, making them suitable for flexible devices.

# 1. Introduction

Virtual substrates are crystalline buffer layers that are epitaxially grown on substrates that differ structurally. Also known as metamorphic buffer layers, these epitaxial layers can address the limitations of available native substrates in three significant ways: (i) by enabling the development of novel microelectronic technologies, (ii) by expanding the application range of existing devices, and (iii) by facilitating entirely new technologies that stabilize material phases with otherwise unattainable properties[1].

However, technical challenges include scalability, high productivity, and compatibility among the various manufacturing steps required to harness these virtual substrates and bring them to market[2].

The successful implementation of epitaxial thin films on silicon (Si) is essential for industrial processing utilizing microelectronic technologies. The silicon substrate must be chemically and structurally compatible with the desired film to prevent the formation of unwanted interfacial defects and to ensure economic feasibility for large-scale production. When the native silicon substrate does not meet these requirements, virtual substrates, such as buffer layers that allow for changes in lattice parameters, structure, or chemistry, can provide a viable solution.

For example, metamorphic buffer layers have succeeded at an industrial scale for silicon-germanium (SiGe), III–V compound semiconductors, and group III-nitrides. In all these instances, metamorphic epitaxial materials have enabled both (i) the use of more cost-effective and larger non-native substrates for electronic devices[4] and (ii) attained record performances of a wide range of applications, including efficient photovoltaics multijunction[5], III–V based lasers on Si[6,7], transistors with high electron mobility[7] and heterojunction tunnel field effect transistors for low-power logic[8].

Given their wide range of outstanding properties, addressing the challenge of materials integration for functional oxides with silicon using virtual substrates presents an opportunity to develop new electronic devices in the More than Moore era[9]. However, the majority of metal oxide phases are deposited and crystallized through high-temperature treatments ranging from 200–800 °C, which are incompatible with standard CMOS technology[10]. Moreover, combining silicon with structurally, thermally, and chemically reactive oxides in hybrid structures poses significant challenges[11]. Consequently, most oxide materials are only available as bulk crystals, and there are currently no reliable and cost-effective processes for depositing them as single crystal thin film at the industrial level. Therefore, these materials can only be configured as devices through direct bulk micromachining or hybrid integration methods, which diminishes the performance of these devices due to the difficulties associated with preparing them directly on a single-crystalline oxide substrate or using bonding techniques[22].

A metamorphic buffer technology for oxides has been developed that integrates $SrTiO_3$ thin films through molecular beam epitaxy (MBE), allowing for integrating other complex cubic perovskite oxides on silicon[12–14]. This technology is continuously evolving in search of an industrial solution for these virtual substrates[15]. While MBE systems have been used on an industrial scale since the 1980s primarily for the growth of high-quality binary semiconductor structures, the increased growth rates have led to the production of nonstoichiometric $SrTiO_3$ films[16]. Also, maintaining a low pressure of the order of $10^{-12}$ Torr during the multielement deposition is cumbersome and more expensive than other physical and chemical deposition processes[3,17].

Several challenges remain: (i) the need for wafer-scale and cost-efficient processes for depositing uniform, high-quality epitaxial functional oxides thin films at atmospheric conditions that can serve as a virtual substrate for integrating other relevant non-perovskites oxides on silicon, (ii) the development of material deposition techniques for thin film form and nanostructures with properties superior to those of bulk materials and (iii) the use of nontoxic, lightweight and abundant elements.

In this context, ZnO appears multifunctional, exhibiting remarkable piezoelectric, thermoelectric, sensing, optical, and catalytic properties. It is abundant, inexpensive, and nontoxic, addressing the aforementioned challenges. Consequently, there is significant interest in developing affordable synthetic methods suitable for the deposition of ZnO over large areas. Indeed, ZnO is an n-type semiconductor oxide with a direct band gap of ~3.3 eV, similar to that of GaN[18]. Doping with Al or Mg shows a significant increase in electrical conductivity while maintaining transparency and chemical and thermal stability[19]. Thus, it has been proposed as a suitable material for fabricating transparent and conducting electrodes in solar cells, a natural replacement for indium tin oxide (ITO) in liquid crystal displays, and anode material for organic light-emitting diodes[18]. Excellent photocatalytic and gas-sensing activity has also been reported[20,21].

In its thermodynamically stable wurtzite phase (P63mc), with a=3.249 Å, and c=5.204 Å, ZnO presents a spontaneous polarization along the [001] direction, influencing its thermo-electrical, optical and catalytic properties. This polar direction consists of the most thermodynamically stable planes; thus, the (0001) mixed-terminated surface is the preferred orientation observed in thin films[22]. This preferential orientation and the piezoelectric properties of ZnO make it an ideal candidate for energy harvesters[23].

However, one significant hurdle in developing ZnO-based multifunctional devices is the integration with silicon. Direct growth of ZnO on Si has proven to be extremely difficult, resulting almost exclusively in polycrystalline and textured films, due to the formation of oxide layers on the Si surface, the reactivity between ZnO and Si, a significant lattice mismatch (15.4 %), and a large mismatch in the thermal expansion (60 %). These issues can only be mitigated by employing metamorphic buffer layer technology. Furthermore, native ZnO substrates are not widely available, and their prices are high for large surface areas. Recent studies have demonstrated the successful integration of (0001) ZnO in both 2D and 1D forms on (111) silicon using various buffer layers such as $Lu_2O_3$, $Sc_3O_3$, $GdO_3$, yttria-stabilized zirconia (YSZ), STO, or GaN. These buffer layers were deposited using techniques such as molecular beam epitaxy, pulsed laser deposition[24], or sputtering[25]. However, due to the high cost and low yield, none of these methods are suitable for industrial applications. Moreover, in all cases, ZnO material was deposited and crystallized at high-temperature treatments in the range of 500–800 °C, which are incompatible with standard CMOS technology.

In this study, we developed an innovative strategy for epitaxial growth of ZnO on silicon at extremely low temperatures (100°C) using α-quartz metamorphic buffer technology. Our approach involved tailoring the nucleation, crystallization, microstructure, and thickness of epitaxial α-quartz films on silicon, achieving sizes up to 4 inches.

We then utilized (100)-oriented α-quartz virtual substrates to accommodate the ZnO unit cell at low temperatures ranging from 70°C to 140°C. This process resulted in the formation of horizontal microwire films with the epitaxial relationship [110] ZnO(110) // [100]*Quartz-α(100) // [100]*Si(100). As a result, we obtained perfectly aligned horizontal ZnO microwire layers that exclusively expose non-polar planes, allowing for precise control over the length, width, and percentage of the ZnO crystal coating at the wafer scale.

We leveraged these epitaxial ZnO nanowire films on quartz buffer layers to fabricate large-scale piezoelectric MEMS devices. Additionally, our microfabrication capabilities on the silicon substrate enabled the preparation of (110) ZnO suspended membranes that are 800 nm thick. These membranes can be transferred to flexible substrates, making them suitable for energy applications.

## 2. Epitaxial α-quartz buffer technology

In this work, we aimed to develop a sustainable method to scale up the integration of ZnO films on silicon in order to manufacture piezoelectric devices, as schematized in **Figure S1**. To this end, we first engineered an α-quartz buffer technology by integrating (100)-oriented epitaxial α-quartz thin film on larger surfaces (up to a 4-inch silicon (100) wafer) without compromising the crystal quality, orientation, homogeneity, and piezoelectric properties of the material. To complete this objective, we used spin-coating, a technique more adapted to the microelectronic format, and a modified version of our previous quartz precursor solution[26].

We improved quartz crystallization as part of the scaling-up of the integration process by studying the effect of surfactants on strontium catalyst salt solubility. Therefore, we achieved higher strontium concentrations by incorporating into the α-quartz precursor sol-gel solution Brij-58, a non-ionic surfactant that increases the solubility of the strontium chloride hexahydrate ($SrCl_2 \cdot 6H_2O$) catalyst while preserving the stability of the precursor solution. We observed that Brij-58 induces a homogeneous distribution of $Sr^{+2}$ devitrifying cations through an evaporation-induced self-assembly (EISA) mechanism[27]. This phenomenon might be produced by the surfactant Brij-58, which can form nanometric micelles capable of absorbing small amounts of water[28]. Therefore, these Brij-58 micelles could form during the EISA process of spin-coated doped-strontium silica layers, trapping small quantities of water and $Sr^{+2}$ ions. As a consequence, the molar ratio of Brij-58 ($R_{Brij}$ = 0.044) proved to stabilize strontium salt concentration up to $R_{Sr}$ = 0.1 into quartz sol-gel precursor solution (see **Table S1**), achieving homogeneous nucleation and optimum crystallization of the epitaxial (100) quartz thin films at the wafer scale as shown in **Figure 1a**.

For α-quartz thin films prepared with an $SrCl_2·6H_2O$ molar ratios between $0.030 \leq R_{Sr} \leq 0.035$, the strontium concentration is not sufficient to devitrify the amorphous sol-gel and silicon native $SiO_2$ layer and thus impeding quartz nucleation and epitaxial crystallization. However, we found that silica layers doped with an $SrCl_2·6H_2O$ molar ratios greater or equal to $R_{Sr} \geq 0,0375$ crystallize uniformly over the entire 2-inch Si wafer substrate (Figure 1). More importantly, with a molar ratio $R_{Sr}$ = 0.1, the epitaxial (100) α-quartz layers reduce their degree of mosaicity, reaching uniform 2-inch (100) quartz/ (100) silicon with a top mosaicity degree of 1.13° (**Figure 1a** and **Figure S2.**).

Spin-coating deposition conditions of sol-gel precursor solution also influence the crystallization of quartz layers. We have carefully examined the impact of centrifugation speed on the thickness of the deposited silica layer and the crystallinity of the final epitaxial quartz thin film. As a result, samples manufactured with a centrifugation speed between $1000 < \omega < 3500$ rpm and a fixed optimized surfactant molar ratio of $R_{Brij}$ = 0.044 and $R_{Sr}$ = 0.1 were prepared (**Figure S3**). We observed that silica films deposited with a centrifugation speed of $\omega$ = 2000 rpm for 30 seconds obtained 220 nm thick uniform silica layers with the best mosaicity degree at the silicon wafer scale after crystallization at 1000 °C under an air atmosphere. Taking advantage of all these results, we were able to fabricate uniform and flat thin films of epitaxial (100) quartz at the wafer scale optimal for MEMS applications with minor thickness variations in the 170 nm to 285 nm range (see **Figure S4**).

Thanks to the optimum conditions for the integration of quartz on Si, we carried out the crystallization of strontium-doped silica layers deposited on 2, 3- and 4-inches silicon wafers. Hence, these results demonstrate the scaling of the integration process. Furthermore, our results showed an equivalent crystallization irrespective of the substrate size, indicating homogeneous nucleation across the entire substrate with mosaicities between 1.5° and 1.8° and free of macroscopic defects (**Figure 1c and Figure S5**).

Pole figures of the Si{100} and α-quartz {101} reflections of **Figure 1d** confirmed the previously observed epitaxial growth and the existence of two well-defined quartz crystal domains that have the same epitaxial relationship of [100]Si(100)//[100]*α-quartz(100)*, in agreement with our previous report[29]

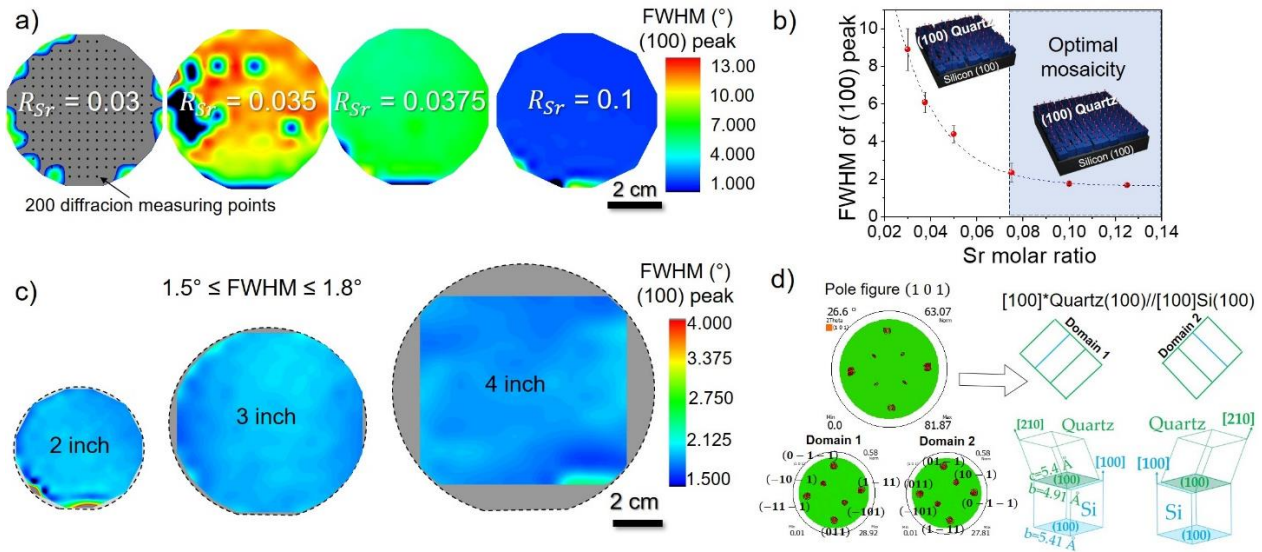

**Figure 1. Wafer-scale integration of epitaxial (100) α-quartz thin films on (100) silicon up to 4 inches**. **a)** Nucleation and crystallization of epitaxial (100) α-quartz layer on a 2-inch wafer as a function of Sr devitrifying agent molar ratio. Mosaicity heat maps around the (100) α-quartz peak indicate homogeneous nucleation from a Sr ratio of 0.0375. Notice that each textural heat

map is composed of 200 diffraction measuring points performed with a collimator of 1 mm$^2$ and a 2D XRD detector in order to perfectly cover the whole silicon wafer surface **b)** Plot showing the decrease in FWHM value around the (100) quartz peak as a function of Sr molar ratio. **c)** Mosaicity mappings around the (100) quartz peak, demonstrating the homogeneity of quartz layers on 2-inch, 3-inch, and 4-inch wafers. **d)** Indexed pole figure of α-quartz {101} reflections that demonstrate the existence of two well-defined quartz crystal domains as schematically represented in 3D on the right side of the figure.

**3. Epitaxial growth of horizontal (110) ZnO microwire films on quartz buffer technology**

Next, we used the wafer scale integration of epitaxial (100) quartz thin films as a new crystalline buffer strategy to integrate epitaxial wurtzite structure ZnO on silicon. We observed the crystallization of fully (110) textured ZnO microwires on a quartz buffer layer by using a controlled low-temperature hydrothermal deposition method at 2.1 bar and 100°C in a water solution containing a Zn salt in the presence of hexamethylenetetramine (HTMA). More preciously, 2D-X-ray diffraction scans and field emission scanning electron microscopy (SEM-FEG) revealed the crystallization of horizontal (110) textured ZnO crystals along the α-quartz buffer layer surface through a non-classical self-assembly mechanism of small ZnO particles as described by Bitenc et al.[30] (see **Figure 2**).

(110) ZnO microwires film precursor solution consisted of a mixture of an aqueous solution of zinc nitrate and hexamethylenetetramine or HMTA was used for the hydrothermal synthesis of epitaxial ZnO microwires on (100) α-quartz virtual substrates. The epitaxial growth of ZnO microwires is conducted in a hermetically sealed autoclave heated at a constant temperature, typically between 80°C and 140°C, for 2 hours as illustrated in Figure 2. Notably, positioning the silicon wafer inside the reactor plays a key role in synthesizing epitaxial horizontal ZnO microwires on quartz virtual substrates by the hydrothermal route. Indeed, we observed that positioning the substrate face down favors a high reproducible growth of horizontal (110) ZnO microwires on the substrate plane, avoiding the presence of ZnO powder on the quartz buffer layer surface (see the synthesis details in the experimental section). Moreover, the quartz virtual substrate must be pre-conditioned by a hydrothermal surface cleaning cycle with HMTA solution at 110°C for one hour to remove impurities and strontium carbonate particles formed during the crystallization of the quartz layer. As a result, a smooth and perfect surface of quartz virtual substrate is obtained for the epitaxial growth of ZnO.

During the hydrothermal growth of textured (110) ZnO microwires films, we also detected the formation of undesirable $Zn_5(OH)_8(NO_3)_2 \cdot 2H_2O$ (ZnNOH) crystalline phase[31] (see **Figure 2**). The presence of ZnHOH large crystals in the ZnO precursor solution can produce inhomogeneities on the final epitaxial (110) ZnO microwire films (see **Figure S6**). Therefore, we used a zinc nitrate and HMTA solution proportion of (2:1) that increases the concentration of $Zn^{+2}$ cations. Consequently, the chemical equilibrium is shifted in the direction of ZnO microwire formation, avoiding the formation of ZnHOH in the initial precursor solution (see **Figure S7**). As a result, we reached the synthesis of homogeneous epitaxial (110) ZnO microwire films on silicon with a horizontal growth, i.e., the longitudinal polar (001) axis parallel to the substrate plane, which is very advantageous for many applications such as high frequency and low loss sensing[32], harvesting energy[33], catalysis[34], or thermoelectricity[35]. Remarkably, no method allows simple access to the self-assembly horizontal epitaxial (110) ZnO microwires on silicon with a unique texture and perfect control over crystal dimensions and crystallinity. Most ZnO integration processes consist of the vertical growth of 1D and 2D ZnO structures along the (001) orientation. Only recent examples have synthesized horizontal (110) ZnO thin films prepared by PLD on SrTiO$_3$[36], AlLaO$_3$[37], or Al$_2$O$_3$[38] single-crystal substrates. In another approach, epitaxial 110 ZnO nanowires were achieved by Au-catalyzed vapor transport[39], although on non-silicon substrates.

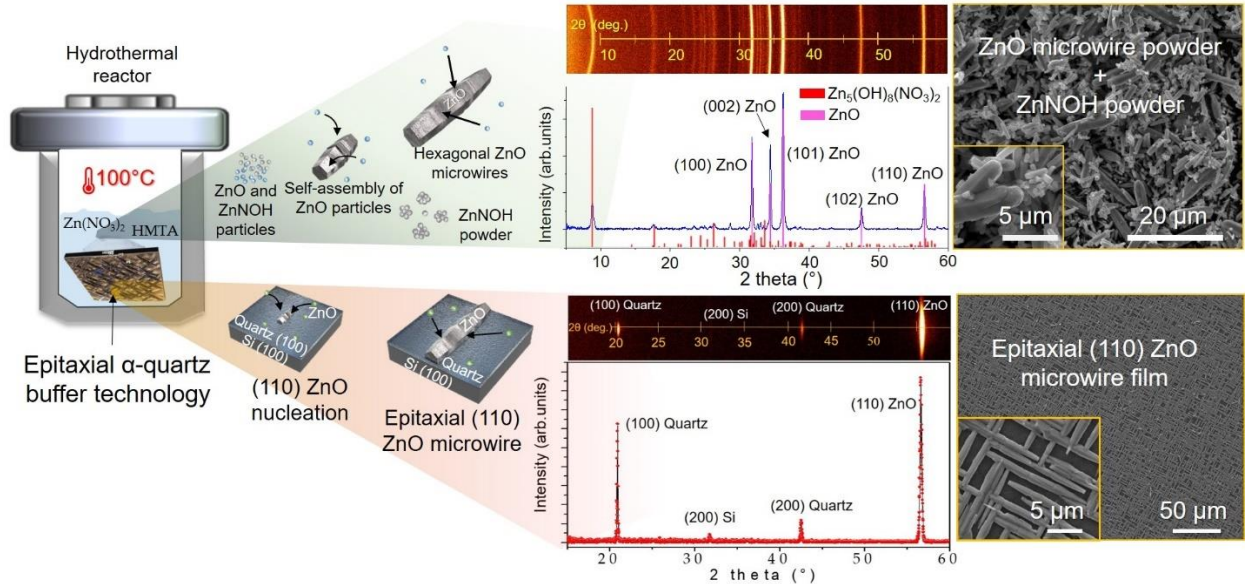

**Figure 2. Hydrothermal growth conditions of epitaxial (110) ZnO microwires on (100) α-quartz virtual substrates.**

Next, we studied the epitaxy of ZnO on α-quartz virtual substrate films with three-dimensional electron diffraction (3D ED) in a cross-section prepared for TEM observation. 3D ED is a technique that enables the reconstruction of the diffraction space from nanometer-sized regions and thus becomes ideal for the structural characterization of the interface[40]. We collected three 3D ED datasets with a 120-nm precessed electron beam corresponding to ZnO, quartz and silicon close to the interface during a single tilt scan of the TEM goniometric stage[41] (**Figure 3 and S**8). The acquired diffraction data was processed to determine the unit cells of ZnO, quartz and silicon and understand the crystallographic relation between them. The parameters found for quartz (a=b=4.9107(3) Å, c=5.4434(4) Å, α=β=90.0°, γ=120.0°) confirmed the presence of the α polymorph and a dynamical refinement allowed the determination and refinement of the corresponding absolute crystal structure (**Table S2**). The orientation matrix found for each crystalline phase also enabled the analysis on how the determined averaged unit cells of the measured areas fit together. The angle between $\vec{c}_{ZnO}$ and $\vec{c}_{quartz}$ is 177.5 ± 0.2° and between $\vec{a}_{ZnO}$ and $\vec{a}_{quartz}$ is 28.0 ± 0.2°, thus the structure of ZnO and α-quartz are slightly tilted with respect to each other at the interface. On the other hand, the angle between $\vec{c}_{quartz}$ and $\vec{c}_{Si}$ is 179.0 ± 0.2° and between $\vec{a}_{quartz}$ and $\vec{b}_{Si}$ is 179.0 ± 0.2°, which indicates that a more orthogonal epitaxial growth occurs between α-quartz and silicon.

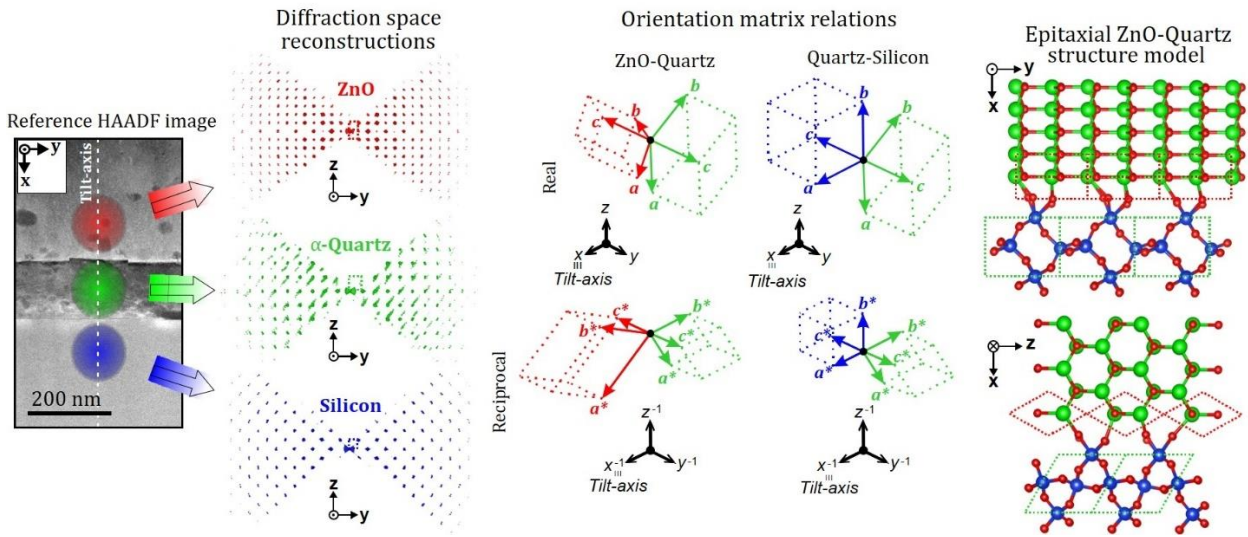

**Figure 3. Structural and epitaxial characterization of (110) ZnO // (100) α-quartz thin films on (100) silicon wafer by means of 3D ED. The reference high-angle annular dark field (HAADF) image shows the areas in red (ZnO), green (α-quartz) and blue (silicon) from where the 3D ED data were acquired. The size of the color-filled areas corresponds to the size of the electron beam (around 120 nm), and the cross-section was inserted in the TEM holder in an orientation that the tilt axis was parallel to the growth direction (as indicated**

**with a dashed-white line). Diffraction space reconstructions and determined unit cells are displayed for each phase with the same color for consistency. The *x-y-z* reference system is the same across all the sub-figures.**

The study of the crystal domain structure of heteroepitaxial (110) ZnO microwire thin films on (100) α-quartz reveals the presence of two well-defined domains, as shown in the pole figure of the ZnO {110} (**Figure 4a**). The ZnO single crystal microwires align their unit cells on the two perpendicular domains of the epitaxial (100) α-quartz thin film, creating a percolated network of perpendicular planar microwires (**Figures 4b and 4c**). High-resolution transmission Electron Microscopy (HRTEM) images and automated crystal phase and orientation mapping in TEM have been utilized to examine further the crystal domain structure of the film at the nanoscale, as depicted in **Figure 4d**. The HRTEM cross-sectional image at the intersection of the two ZnO microwire crystal domains shows an epitaxial junction characterized by an in-plane orientation where (1-10) ZnO $_{domain\ 1}$ [1-10] is parallel to (001) ZnO $_{domain\ 2}$ [001] and an out-of-plane orientation where (110) ZnO $_{domain\ 1}$ [110] is parallel to (110) ZnO $_{domain\ 2}$ [110]. The phase and orientation maps in Figure 4d provide a direct method for acquiring quantitative information about the microwire orientation at the nanoscale. Cross-sectional phase and orientation maps of a ZnO microwire single crystal (domain 2) demonstrate a perfect (110) out-of-plane orientation of the ZnO microwires on the (100) α-quartz domains, without structural or chemical defects, secondary phases, and featuring sharp interfaces between the different layers. This observation is consistent with the 2XRD, HRTEM, and 3D electron diffraction analysis findings. **Figure 4f** illustrates a linear correlation between the mosaicity value around the (110) orientation of the ZnO microwires and the (100) reflection of the quartz buffer layer. This significant result enables the tuning of the mosaicity of the ZnO microwire layer through the quartz buffer layer, achieving a top mosaicity degree of 2.5°.

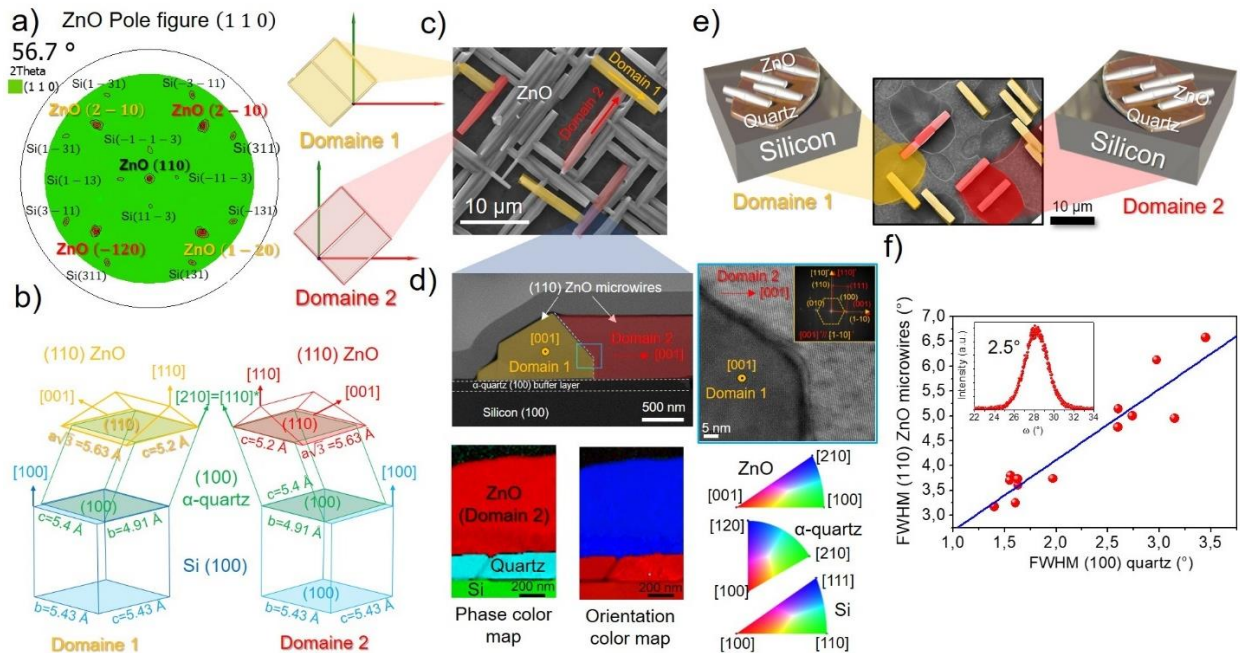

**Figure 4. Crystal domains structure characterization of heteroepitaxial (100) α-quartz // (110) ZnO thin films on (100) silicon wafer**. a) Indexed pole figure of ZnO {110} reflections that demonstrate the existence of two well-defined ZnO crystal domains. b) 3D schematics of (110) ZnO // (100) α-quartz // (100) Silicon unit cells in the two crystal domains. c) SEMFEG image of ZnO epitaxial microwires on quartz virtual substrate highlighting the two existent crystal domains. (d) TEM and HRTEM images of ZnO crystal domains. The Fourier Transformation (FT) from the areas marked in blue is included as an inset and shows the epitaxial relationship between the ZnO two domains crystals. Below are the phase and orientation maps obtained from 4D-STEM datasets; the red color refers to the ZnO phase (domain 2), the light blue to α-quartz and the green to silicon. The colormaps are for the orientation map, which indicate the colors related to all possible symmetrically independent directions according to the Laue class of the different phases. (e) 3D

schematics and FEGSEM image of ZnO crystal domains on quartz virtual substrate. (f) Control of the mosaicity value of epitaxial (110) ZnO microwires through the (100) α-quartz buffer layer. The plot shows the FWHM value of the (110) ZnO peak dependence with the FWHM value of the (100) α-quartz.

Understanding and controlling the out-of-plane crystalline strain behavior ($\varepsilon_{yy}$) is crucial for optimizing the piezoelectric properties of heteroepitaxial thin films. The out-of-plane strain $\varepsilon_{yy}$ significantly influences the piezoelectric performance of these films, as demonstrated by various studies[42]. Therefore, the relationships between strain and properties facilitate the precise engineering of piezoelectric thin films, enabling the development of more efficient and tailored devices for numerous applications, including electronics, sensors, and actuators.

To explore this phenomenon, we investigated the crystalline strain in epitaxial (110) ZnO microwire films deposited on α-quartz (100) virtual substrates by acquiring four-dimensional (4D) STEM diffraction maps (see **Figure 5, Figure S9 and Figure S10** and further technical details in the supplementary information section). We found that in both epitaxial (110) ZnO crystal domains—domain 1 and domain 2—the out-of-plane epitaxial strain decreases from a strained region close to the α-quartz virtual substrate to a fully relaxed zone and then transitions to a tensile strain as the thickness of the ZnO microwires increases (see **Figure 5a**).

Additionally, the strain analysis presented in Figure 5a reveals that the $\varepsilon_{yy}$ values vary significantly between different (110) ZnO crystals, with some regions exhibiting up to 2.5 times higher strain. The boundaries between ZnO microwires with differing crystalline domains exhibit a compressive strain of $\varepsilon_{yy} = -1\%$. To assess the impact of this strain structure on piezoelectric properties, we conducted piezoelectric force microscopy imaging on epitaxial (110) ZnO with varying crystal domains and compared the results with the 4D STEM diffraction maps.

Our observations revealed a complex piezoelectric response at the micrometer scale, where the PFM amplitude and phase maps indicate a distinct piezoelectric response from two different ZnO microcrystals domains (see **Figure 5c and d**). This AFM tip amplitude variation (see histograms from **Figure 5 e**) might be attributed to the differing strain values identified in the 4D STEM diffraction maps of two ZnO microcrystals with different crystalline domains. Furthermore, we noted an increased piezoelectric response at the grain boundaries between ZnO microwires (see Figure 5c), which correlates with the compressive strain $\varepsilon_{yy} = -1\%$ of a ZnO microwire with domain 1 in Figure 5a.

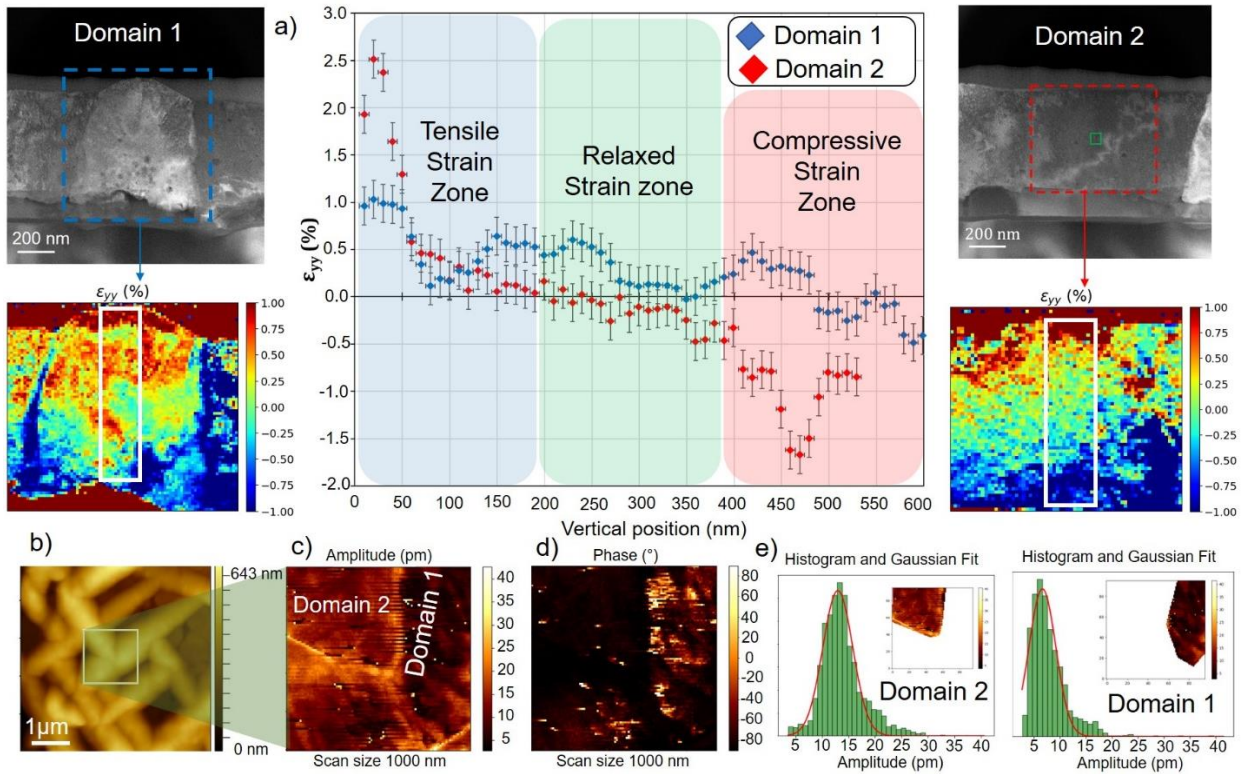

**Figure 5. Strain correlated piezoelectric analysis at the microscale of (110) ZnO microwires films.** (a) Four-dimensional (4D-)STEM diffraction maps acquired to quantify and visualize the crystalline strain of the two crystalline domains of epitaxial (110) ZnO films. (b) Topographical AFM image of a (110) ZnO epitaxial film on quartz virtual substrate. (c) PFM amplitude image of two ZnO crystal domains, (d) Phase map and (e) Amplitude histograms of the PFM amplitude in the regions "domain 2" and "domain 1" in figure (c).

Thanks to the combination of these approaches and considering the interplay between strain, microstructure, and PFM images, we established a correlation between crystalline strain and piezoelectric response in epitaxial (110) ZnO at the microscale. Importantly, from these data, we can confirm a complex piezoelectric response at the microscale from epitaxial (110) ZnO films which could benefits piezoelectric devices.

Next, we investigated the nucleation and crystal growth mechanism of epitaxial (110) ZnO microwires by modifying the hydrothermal synthesis conditions. This approach allowed us to gain precise control over the size and length of the (110) ZnO single-crystal microwires, paving the way for scaling up the fabrication process. We examined the influence of hydrothermal synthesis time on the crystal growth and nucleation of epitaxial ZnO (110) microwires, producing samples that were synthesized for durations ranging from 15 to 300 minutes at 100°C, using similar α-quartz buffer layers (see **Figure S11**). SEM-FEG images indicated a progressive growth of (110) ZnO microwires through the self-assembly mechanism of small ZnO particles[30]. As a result, the microwires measured between 1 and 13 micrometers in length, with widths ranging from 0.5 to 1.5 micrometers (see **Figure S11a** and **S11b**). We observed logistic growth patterns in both the length and width of the ZnO microwires, reaching saturation levels of approximately 13 µm in length and 1.5 µm in width after 120 minutes of synthesis (see **Figure S11c** and **S11d**).

We studied the effect of temperature on the hydrothermal synthesis of epitaxial (110) ZnO microwires. To do this, we prepared samples using hydrothermal synthesis temperatures ranging from 70°C to 140°C, using similar quartz buffer layers and a synthesis time of 120 min. **Figure 6a** presents SEM images of the epitaxial (110) ZnO microwire samples, illustrating a gradual increase in the area of the quartz buffer layer occupied by the ZnO microwires. By 140°C, the microwires almost completely cover the surface. We noted a significant contrast between the quartz buffer layer and the ZnO microwires in optical images. This contrast allowed us to efficiently select the surface of the ZnO microwires using image processing software (see **Figure S12**). Consequently, we were able to quantify the surface area occupied by epitaxial (110) ZnO microwires in samples prepared at synthesis temperatures between 70°C and 140°C. The graphic in **Figure 6b** shows that the surface area occupied by ZnO microwires increases exponentially with temperature, covering the entire quartz buffer layer and forming a continuous layer of epitaxial ZnO microwires. These results presented in this section provide insights into how to control the size of ZnO (110) epitaxial microwires and the surface area they occupy on the quartz buffer layer.

Additionally, we have established the temperature and time parameters for the hydrothermal synthesis used to create dense layers of ZnO (110) epitaxial microwires on silicon. This information is particularly valuable for scaling up the manufacturing process of these ZnO (110) microwires, which represents significant progress toward implementing this novel architecture in different device applications. Therefore, and thanks to the optimum conditions for integrating α-quartz buffer layers and ZnO on Si, we crystallized dense 800 nm thick (110) ZnO microwires on 2-inch silicon wafers. Hence, these results demonstrate the scaling of the integration process. Furthermore, our results showed an equivalent crystallization irrespective of the substrate size, indicating homogeneous nucleation across the entire substrate with mosaicities between 2° and 3° and free of macroscopic defects (**Figure 6c**).

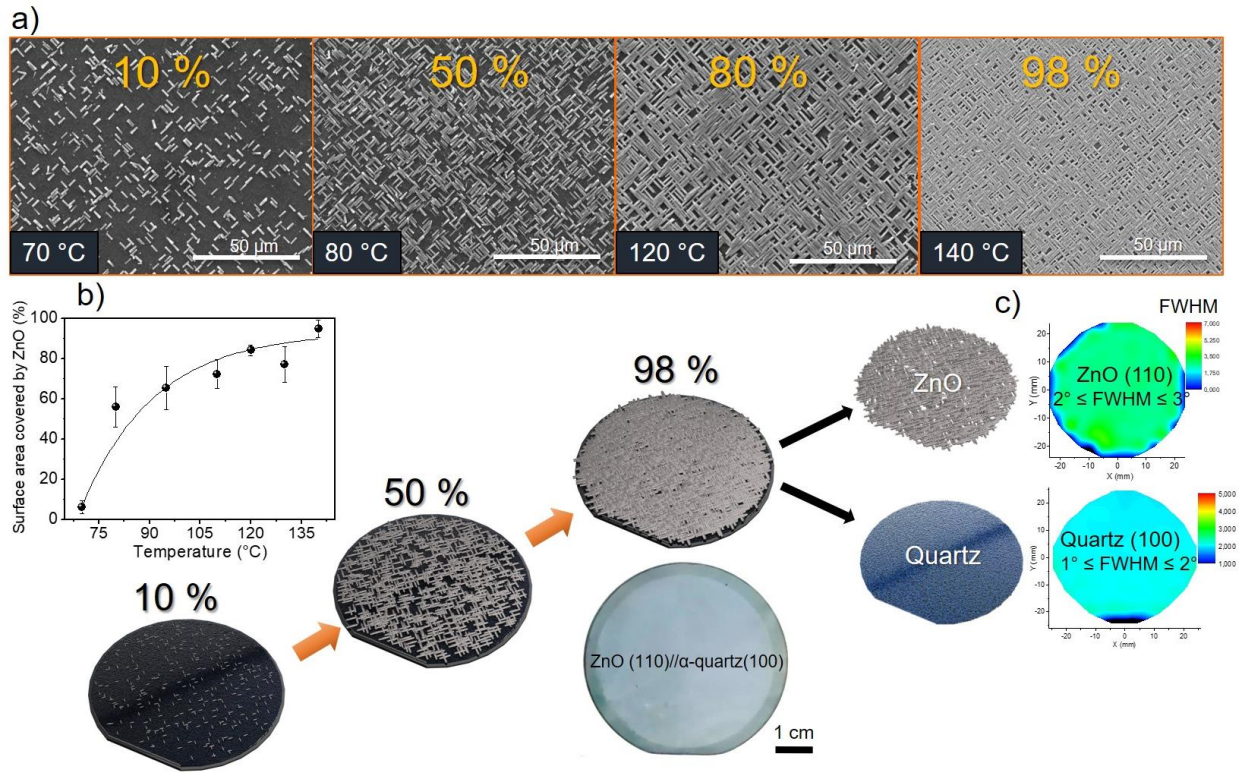

**Figure 6. Wafer-scale integration of epitaxial (110) ZnO microwire thin films on (100) quartz virtual substrates up to 2 inches with controlled crystal size, length, and density**. (a) Top-view SEMFEG images of epitaxial (110) ZnO microwire films obtained through hydrothermal synthesis at different temperatures, demonstrating a 98% microwire film uniformity and homogeneous samples grown at 140 °C up to 2 inches. (b) An exponential saturation plot was obtained from analyzing the surface area occupied by epitaxial ZnO (110) microwires in samples synthesized at different temperatures. Below are 3D schematics and optical images showing the evolution of (110) ZnO microwire films grown at different temperatures. (c) Heat maps of the FHWM values of 800 nm thick epitaxial (110) ZnO microwire films and 120 nm thick epitaxial (100) α-quartz

thin film rocking curves. Therefore, the scale-up of (110) ZnO integration on silicon is demonstrated. Notice that each textural heat map comprises 200 diffraction measuring points performed with a collimator of 1 mm$^2$ and a 2D XRD detector to perfectly cover the whole silicon wafer surface.

To achieve a more significant and controlled thickness of epitaxial ZnO microwire films at a wafer scale, we have developed a multicycle deposition approach by using hydrothermal synthesis (**Figure 7**). This methodology consists of depositing several ZnO cycles (see more details in the experimental section). After the ZnO crystallization process, this multicycle deposition method showed a linear increase of the layer thicknesses up to 2.7 µm (see SEMFEG images in **Figure 5a**) while maintaining the crystal quality, mosaicity, and coating uniformity without the appearance of secondary phases or undesirable ZnO orientations (**Figure 7b-d**). This repetition underscores the scalability of the process, allowing for the production of ZnO microwire films of varying thicknesses to suit different applications.

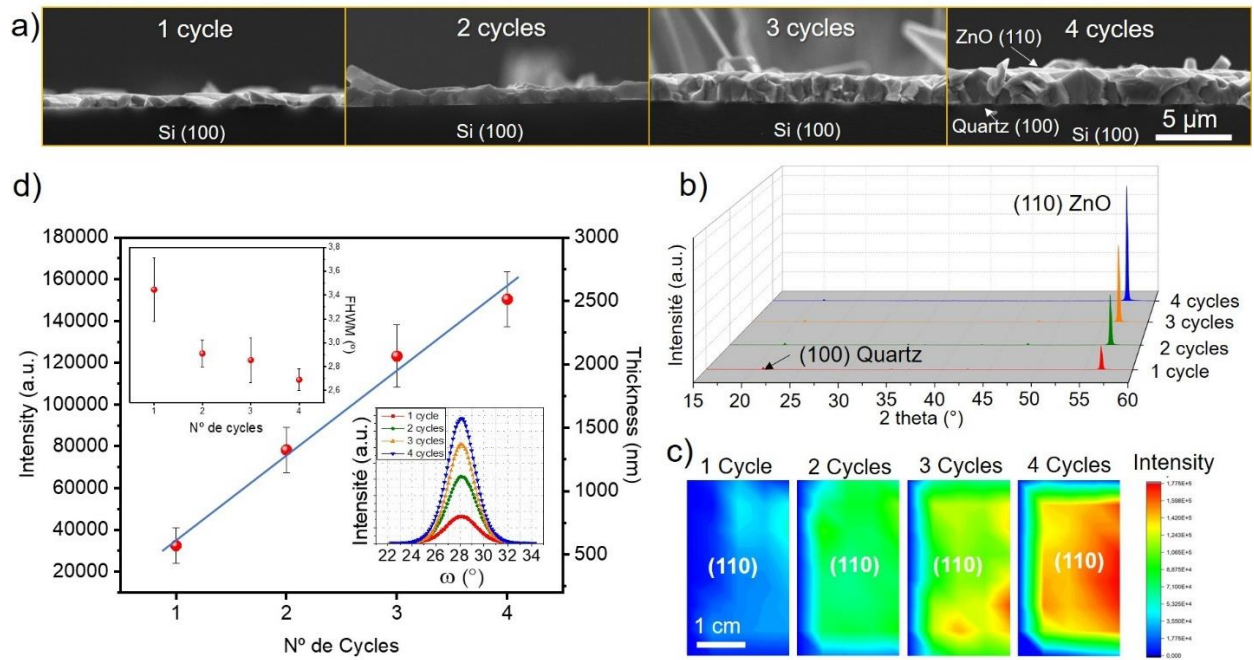

**Figure 7. Control of large-scale epitaxial (110) ZnO microwire film thickness using low-temperature multicycle deposition methodology.** (a) Cross-sectional SEMFEG images of epitaxial (110) ZnO microwire films obtained through successive cycles of deposition using low-temperature hydrothermal synthesis. (b) X-ray diffraction characterization of different epitaxial (100) ZnO microwires films, obtained through several successive hydrothermal synthesis cycles. (c) 2D (110) diffraction ZnO peak intensity maps from samples produced with different numbers of hydrothermal synthesis cycles. (d) Plot showing linear dependence of the (110) crystallographic ZnO reflection intensity values and its corresponding thicknesses as a function of the number of hydrothermal cycles. Insets show the evolution of the rocking curves and FWHM values of the (110) ZnO peak as a function of the number of cycles.

## 4. Epitaxial (110) ZnO based-MEMS fabrication

Integrating high-quality epitaxial functional oxide films and nanostructures on silicon can boost the fabrication of many devices with the traditional Si-based complementary metal-oxide-semiconductor (CMOS) technology[3]. Moreover, advances in micro and nanofabrication technologies open the possibility of implementing a large-scale integration of miniaturized piezoelectric materials into innovative electromechanical devices with nanosized moving parts (MEMS / NEMS) with prospective applications in electronics, biology, and medicine[4-7]. Consequently, by taking advantage of the epitaxial integration of (110) ZnO microwires on silicon combined with both (i) the control of unique physical, chemical, and microstructural properties of (110) ZnO microwires and (ii) advanced microfabrication processes, we scaled up different piezoelectric MEMS as model systems devices (see **Figure 6**).

We developed an innovative microfabrication approach that combines a direct laser engraving process with a selective anisotropic wet etching with tetramethylammonium hydroxide (TMAH) from the backside of the silicon substrate. The device's top face, i.e., the active (110) ZnO microwire film, is protected by screw-tightened wafer holders (also known as wafer chucks) (see **Figure S13**). To protect the backside silicon wafer from the chemical etching, we developed an innovative layered system of protective coatings made 400 nm thick SiN /400 nm thick $SiO_2$/ 400 nm thick SiN which made it possible to resist the chemical attack of TMAH for more than 20 hours at 84 °C. (see **Figure S14**).

Thanks to an LPKF ProtoLaser S4 machine, we can control the laser etching process with high precision and finish the fabrication of the membrane with chemical means (see **Figure S15**). This microfabrication process is not only capable of producing membranes with controlled shapes and thicknesses by reducing chemical etching time on thicker and cheaper standard silicon (100) substrates but also cantilevers, bridges, or more complex resonator structures by a final laser layout of the device. These microfabrication capabilities open the possibility of engineering microelectromechanical sensor devices of epitaxial (110) ZnO on silicon wafers with high mechanical susceptibility, efficient piezoelectric transduction, and low sensitivity to environmental variations. The possibility of engineering epitaxial horizontal (110) ZnO microwire films on silicon in the form of thin membranes with thickness between 1 µm and 4 µm might produce high mechanical deformability and, thus, efficient piezoelectric transduction that is expected for these materials owing to its high electromechanical coupling factor[43]. In addition, the control of ZnO microwire aspect ratio ordered uniformity and doping level opens the possibility of producing more efficient devices. This is supported by the fact that ZnO micro and nanostructures have generated considerable output power, mainly due to their elastic structure, mechanical stability, out-of-plane polarisation, and appropriate bandgap[44]. **Figure 8a** and **Figure 8b** exhibit the first devices on the microfabrication of a 2µm thick and 25 mm² surface piezoelectric (110) ZnO membrane resonators on doped silicon (100). These membranes have a high mechanical quality factor (> 1600) in the air (See **Figure S16**) and can be efficiently activated by their movement from the linear converse piezoelectric effect (**Figure 8a**). Moreover, the (110) ZnO microwire film can be easily patterned, allowing for the fabrication of piezoelectric patterned (110) ZnO microwire membranes with different designs such as micrometric serpentines or lines (**Figure 8b**) (see experimental part for more details).

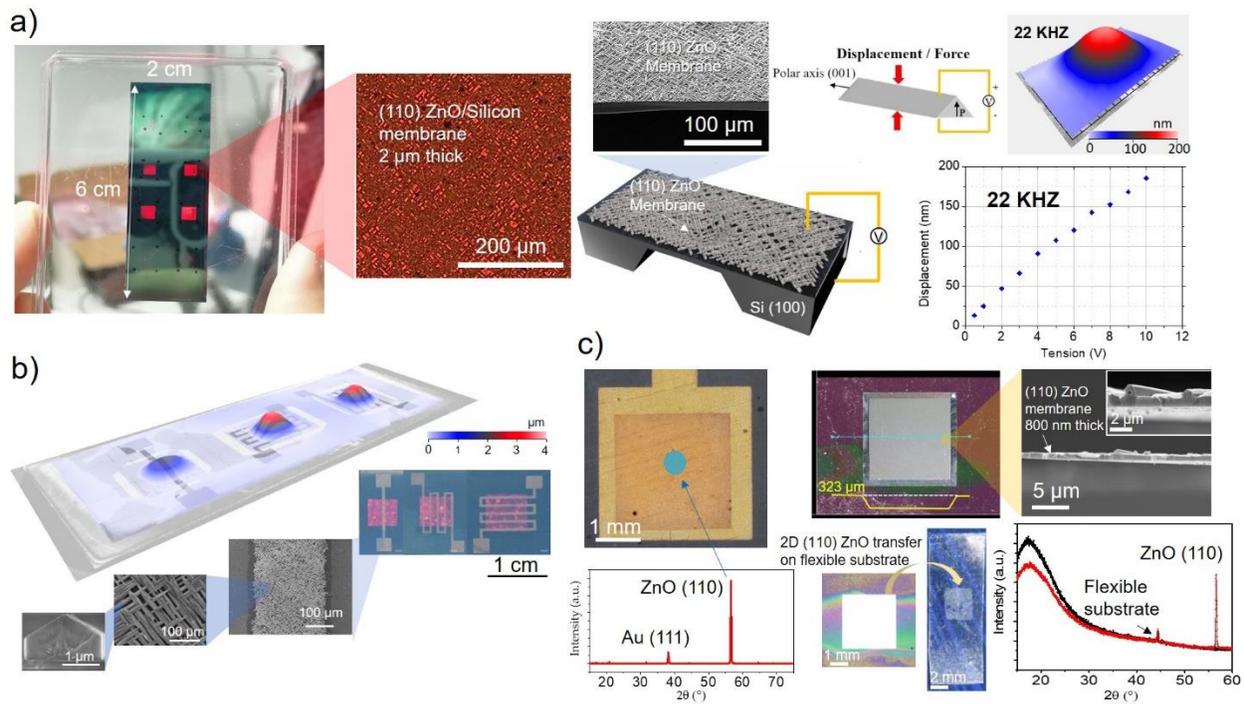

**Figure 8. Cost-efficient and large-scale microfabrication processes developed in piezoelectric ZnO membranes and other resonating structures.** (a) microfabrication of a 2µm thick and 25 mm² surface piezoelectric (110) ZnO membrane resonators on doped silicon (100). Moving from the linear converse piezoelectric effect can efficiently activate these membranes. (b)

Microfabrication of piezoelectric patterned (110) ZnO microwires membranes with different designs such as micrometric serpentines or lines. These membranes can vibrate at the micrometric scale. SEMFEG and optical images show the (110) ZnO microwires patterned films. (c) Micro and nanofabrication of pure (110) ZnO microwire-based resonators by removing the quartz buffer layer and silicon substrate. Cross-sectional SEMFEG images of the 800 nm thick (110)-ZnO membrane show the absence of the α-quartz buffer layer. XRD diffractogram on the (110) ZnO membrane confirms the total etching of the quartz buffer layer and the silicon substrate. These membranes can be easily transferred to a flexible substrate, as shown in the XRD diffraction pattern from the optical image sample in the right part of the figure.

We explored the possibility of fabricating pure (110) ZnO microwire-based MEMS resonators by removing the quartz buffer layer and silicon substrate. Therefore, this paved the way for transferring (110) ZnO membranes into flexible substrates for planar (110) ZnO nanogenerators. To this aim, we fabricated an 800 nm thick (110) membrane of 9 mm$^2$ by completely removing the quartz buffer layer and silicon substrate (see **Figure 8c**). To fabricate this resonator type, we developed a modified version of our previous etching process (see **Figure S17**) to stabilize the (110) ZnO membrane over large surfaces. In that case, we used a poly(methyl methacrylate) PMMA transparent thin film to both electrically isolate the top electrode and mechanically stabilize the thin (110) ZnO membrane. To remove the silicon substrate, we changed the etching agent from TMAH to KOH, allowing us to make the process more environmentally friendly, safer, and faster and avoiding the need to use a clean room. Finally, this (110) ZnO microwire membrane can be easily transferred to flexible substrates, as shown in Figure 6c. (see experimental section).

These results are promising for the realization of sensors and sustainable wafer-scale integrated ZnO sensor devices with piezoelectric transduction that can operate without batteries, powered by mechanical energy harvested from different energy sources, including mechanical vibrations or photoacoustic forces.

**Conclusions**

In conclusion, we have achieved a cost-effective and wafer-scale integration of (100) α-quartz thin films on Si (100) using chemical methods, addressing a long-standing challenge in synthesizing high-quality quartz thin films on an industrial scale. Through chemical processes and thermal treatment, we have effectively controlled the nucleation, crystallization, microstructure, and thickness of epitaxial α-quartz films on silicon up to 4 inches while maintaining a coherent crystalline interface between (100) quartz and (100) silicon.

As a result, we employed (100) α-quartz wafer-scale bottom-up integration as a novel buffer strategy to integrate epitaxial wurtzite structure ZnO on silicon. We demonstrated that ZnO can accommodate its unit cell on the two perpendicular domains of epitaxial (100) α-quartz thin films at extremely low temperatures (between 70°C and 140°C), producing horizontal microwire films with a unique epitaxial relationship of [110]ZnO(110) // [100]*Quartz-α(100) // (100) Silicon.

This chemical process is simple, inexpensive, and quick to implement. Additionally, it allows for the production of wafer scale epitaxial (110) ZnO microwire layers up to 2 inches with high yields and precise control over the length, width, and percentage of coating of ZnO crystals.

Thanks to the combination and correlation of 4D-STEM electron diffraction and PFM advanced techniques, we established a correlation between crystalline strain and piezoelectric response in epitaxial (110) ZnO at the microscale. Importantly, from these data, we can confirm a complex piezoelectric response at the microscale.

Consequently, we were able to fabricate sustainable large-scale piezoelectric (110) ZnO MEMS and other resonating structures by developing innovative microfabrication processes. We created the first 800 nm thick single-crystal (110) ZnO microwire membranes that can be transferred onto a flexible substrate. Collectively, this work paves the way for low-cost and miniaturized single-chip solutions for piezoelectric devices and has potential applications in various sectors of electronics and sensing devices.

**Materials and Methods**

**Synthesis of silica sol-gel thin films**

*Sol-gel preparation and wafer scale deposition and growth of epitaxial (100) quartz on silicon*

All chemicals were obtained from Sigma-Aldrich and used as received without further purification. Solution A was prepared by dissolving 1 g of Brij-58 (CAS number 64-17-5) in 23.26 g of absolute ethanol (CAS number 9004-95-9), followed by the addition of 1.5 g of HCl (37%) (CAS number 7647-01-0) and 4.22 g of tetraethyl orthosilicate (TEOS) (CAS number 78-10-4). The solution was stirred for 18 hours. A 2M aqueous solution of $Sr^{2+}$ (solution B) was prepared using $SrCl_2 \cdot 6H_2O$ (CAS number 10025-70-4). Solution C, used for the deposition of Sr-silica films via spin coating, was prepared by adding 275 µL of solution B to 10 mL of solution A, followed by stirring for 10 minutes. Films were deposited within 40 minutes of preparing solution C to avoid the instability of $Sr^{2+}$. The $Sr/SiO_2$ molar ratio in solution C was 0.05, with a final molar composition of TEOS:Brij-58:HCl:EtOH: $SrCl_2$ = 1:0.44:0.7:25:0.1.

Gel films on Si(100) substrates (until 4 inches) were prepared using a vacuum-free Ossila spin coater. During the dip-coating process, the ambient temperature and relative humidity were set to 25 °C and 40%, respectively, and the deposition conditions were set at a spinning speed of 2000 rpm for 30 seconds. After deposition, films were consolidated with a thermal treatment at 450 °C for 5 minutes in an air atmosphere. The multilayer gel films were obtained by repeating the process of monolayer preparation several times on the same substrate.

As-prepared gel films were introduced into a furnace already at 1000 °C in an air atmosphere and held at this temperature for 300 min. The crystallized films were recovered after cooling down the furnace to room temperature.

*Synthesis and growth of epitaxial (110) ZnO nanowires films on quartz virtual substrates.*

The epitaxial growth of ZnO (110) microwire films on (100) α-quartz virtual substrates can be divided into three steps:

1. Preparation of the ZnO chemical precursor solution.
2. Cleaning cycle of the (100) α-quartz pseudosubstrate.
3. Epitaxial growth of ZnO microwires on the α-quartz pseudosubstrate using hydrothermal synthesis.

In step 1, a ZnO precursor solution (Sol F) was prepared the day before the epitaxial growth (step 3) by mixing two volumes of a 0.1M aqueous solution of hexahydrate zinc nitrate ($Zn(NO_3)_2·6H_2O$) (Sol D) with one volume of an 0.05 M aqueous solution of hexamethylenetetramine hexahydrate (($CH2)_6N_4·6H_2O$) (HMTA) at a concentration (Sol E) (see **Figure S7**). Notice that this mixture needs to be stirred for a minimum of 24 hours. Step 2 involves cleaning the quartz/Si substrate by placing it inside the autoclave and ensuring the epitaxial quartz layer faces downwards. Then, 25% of the autoclave volume is filled with the HMTA solution (Sol E). The hermetically sealed autoclave is then heated in an oven at 110°C for one hour. Next, the substrate is carefully removed and rinsed with ultrapure water and ethanol to avoid damage. This step assists in removing $SrCO_3$ particles generated during the α-quartz layer crystallization, thus preparing a smooth and perfect surface for the epitaxial growth of ZnO. Finally, the epitaxial growth of ZnO microwires is conducted in a hermetically sealed autoclave heated at a constant temperature, typically between 80°C and 130°C, for 2 hours. After this growth cycle, the substrate is removed and thoroughly rinsed with water, followed by an ethanol rinse to eliminate any dust particles produced during synthesis. The epitaxial growth step, a key part of the process, was repeated consecutively to achieve thicker (110) ZnO microwire layers. This repetition underscores the scalability of the process, allowing for the production of ZnO microwire films of varying thicknesses to suit different applications.

*Electron, X-ray diffraction, and optical analysis of epitaxial quartz wafers and ZnO microwire films*

A JEOL F200 ColdFEG transmission electron microscope (TEM) operated at 200kV was used for the structural characterization of the interface via electron diffraction. A Focused Ion Beam (FIB) lamella prepared in a ThermoFisher Scios 2 was inserted in a JEOL analytical tomography holder for three-dimensional acquisitions. The TEM was used in STEM mode, and STEM images (512 x 512 pixels) were recorded using a JEOL high-angle annular dark-field (HAADF) detector in the Gatan Digital Micrograph. Electron diffraction patterns were acquired with a Gatan OneView camera, a CMOS-based and optical fibre-coupled detector of 4096 by 4096 pixels. The electron beam in STEM mode was aligned following a self-made protocol[45] to fulfill the most quasi-parallel condition with a 120-nm beam diameter (Probe size 8 and 10-µm condenser aperture). The acquisition of the 3D ED data that enabled the reconstruction of the diffraction spaces was carried out with the (S)TEM-ADT module[41] a self-developed Digital Micrograph plug-in freely available at https://github.com/sergiPlana/TEMEDtools. This software automatizes the acquisition of 3D ED datasets by automatically inserting and retracting the HAADF detector, setting the diffraction and imaging conditions of the electron beam, and collecting the STEM reference images and diffraction patterns. The program follows the methodology initially implemented but is modified to allow the acquisition of more than one diffraction pattern per tilt angle at any desired positions of the reference image [47]. In this work, three patterns were acquired per tilt angle (at ZnO, quartz and silicon) for an angular range between -40° and 40° of the alpha-tilt angle of the sample holder with a tilt step of 1° (243 diffraction patterns in total). The nominal camera length was set to 500 mm, and a precession angle of 0.76° was used for each diffraction pattern through a P2010 precession unit by Nanomegas SPRL. PETS2 software[47] was used for unit cell determination and data reduction, and Jana2020[48] was used for dynamic refinements through the dyngo module[49].

4D-STEM datasets were also acquired with the same JEOL F200 TEM operated at 200 kV in STEM mode. Probe size 6 and a 10-μm condenser aperture were selected to produce an electron probe of 1 mrad of convergence angle following the reported alignment method[45]. 0.65° of precession was coupled to this beam setting. The 4D-STEM maps were acquired through the synchronization of the DigiScan 3 external scanning unit with the Gatan OneView camera via the STEMx system provided by Gatan Inc. Gatan Digital Micrograph program was used for hardware control. Self-developed Digital Micrograph scripts were used to correct small pattern shifts produced when scanning the beam along large distances. Phase and orientation maps were obtained by using the ASTAR commercial software provided by NanoMEGAS SPRL. Strain maps were processed through custom-made Python scripts *via* functions of the py4DSTEM package[50].

A Nion UltraSTEM operated at 100 kV acquired bright-field images in STEM mode.

Atomic STM resolution images were performed by using a FEI Titan3 operated at 80 kV and equipped with a superTwin® objective lens and a CETCOR Cs-objective corrector from CEOS Company.

The crystalline textures, rocking curve measurements, and epitaxial relationship of quartz films and cantilevers were performed on a Bruker D8 Discover diffractometer equipped with a 2D X-ray detector (3s acquisition each 0.02° in Bragg–Brentano geometry, with a radiation wavelength of 0.154056 nm). 2D and 3D optical images of quartz films and membrane resonators were obtained in a Keyence® VHX7000 optical digital microscope. The microstructures of the quartz films and MEMS were investigated with a FEG–SEM model Su-70 Hitachi, equipped with an EDX detector X-max 50 mm² from Oxford Instruments. AFM studied the topography of quartz films in a Park Systems NX-Scanning Probe Microscopy unit.

*Piezoresponse Force Microscopy characterization*

The Piezoresponse Force Microscopy (PFM) measurements were made on an AFM Brucker Icon with PtSi-NCH probes (from Brucker). The PFM parameters were set at the frequency of 12kHz and with a 5V AC amplitude. This low frequency was chosen to prevent parasitic effects that could amplify the piezoelectric response amplitude[51]. The initial protocol used to evaluate the correct deflection sensibility of the tip consisted in measuring the effective piezoelectric coefficient $d_{33}^{eff}$. value of a flat reference sample with periodically-poled lithium niobate lines (estimated at 7.5pm/V). This coefficient was calculated by dividing the piezoelectric amplitude by the amplitude of the applied voltage. This reference sample scan was made in contact mode. Following this reference verification, scanning of the sample of interests was made in DataCube, the scanned area is divided in pixels and for each pixels the tip approaches the surface, stay in contact while performing PFM (with the above-mentioned parameters of voltage and frequency) and then withdraw before passing to the next pixel. All the measurements were made with the same probe and in ambient atmosphere.

*Four-dimensional (4D-)STEM diffraction maps for strain analysis of epitaxial (110) ZnO film*

Four-dimensional (4D-)STEM diffraction maps were acquired to map and visualize the crystalline strain of the (110) ZnO films. A JEOL F200 TEM ColdFEG operated at 200kV was used to acquire such diffraction data. A quasi-parallel beam was configured in STEM mode with probe size 6 and a 10-µm condenser aperture to produce an electron probe of 1 mrad of convergence angle following an alignment routine available elsewhere[45]. A precession angle of 0.6° was applied to the electron beam via the P2010 signal unit provided by NanoMEGAS SPRL. The 4D-STEM maps were acquired by scanning the electron probe and storing a diffraction pattern in each pixel. A Gatan OneView camera, a CMOS-based and optical fibre-coupled detector of 4096 by 4096 pixels, was employed to collect these patterns, and the STEMx unit was used to synchronize the raster of the beam via the Gatan Digiscan3 scanning unit and the OneView camera. 4D-STEM data were processed through Python scripts (available at github.com/sergiPlana/TEMEDtools) using the py4DSTEM package to generate the strain maps[52].

*ZnO/Quartz/silicon membrane MEMS manufacturing*

A 2-inch quartz virtual wafer coated with the ZnO piezoelectric epitaxial microwire film. The wafer was opened using a laser to create its moving part. The power and repetition rate varies according to the thickness of the wafer and the desired depth of the moving part with different morphologies and sizes. The 2-inch silicon wafer is then placed in support from AMMT to protect the thin film of quartz before being immersed in a TMAH bath heated to 80°C by a water bath. A protective layer is applied by sputtering (PECVD) to the wafer surface, but on the other side from the part covered with ZnO microwires film. This protective layer is composed of three successive depositions: silicon nitride (400nm), silicon dioxide (400 nm), and again silicon nitride (400 nm). Then, the protective layer is etched until the silicon bulk is revealed ether by ICP-RIE with a $CHF_3/O_2$ plasma or with the laser etching system. If the ICP-RIE is chosen for the etching of the protective layer, a lithography step before the RIE etching is mandatory to have the desired membrane design. Then, a laser beam of 3.75 μm in diameter and 5 W in power is used to etch the membrane by uploading a designed membrane file in a laser machine. Due to this flexible and easy etching method, we can etch any shape of membrane we want in a few minutes. Before the TMAH (CAS number 75-59-2) etching step, the native oxide layer that forms on the silicon surface is removed using hydrofluoric acid (BOE 10%, CAS number 76664-39-3).

**(110) ZnO membrane NEMS manufacturing and its transfer on a flexible substrate**

To obtain a (110) ZnO membrane, first, samples are cleaned in acetone/ethanol, then several successive layers of PE-CVD consisting of $Si_3N_4/SiO_2/Si_3N_4$ (500 nm per layer) are deposited on the back side to create an etching mask. The lithography is performed on the protective layers in order to achieve the desired design for the membrane. An Az1518 photoresist is spin-coated at 3000 rpm for 30 seconds and then baked at 110°C for 1 minute. A 20s exposure is done by using a UV lamp through a "stencil" type mask, and the sample is dipped into AZ726 developer for 1 min. The protection is then etched by plasma ICP-RIE $CHF_3/O_2$ to create the openings until the silicon is exposed. The silicon is then chemically etched (KOH 45%) from the back to the quartz underneath the layer. A made-to-measure shield protects the front face. The quartz layer is composed of $SiO_2$, so it acts as a barrier layer for the KOH etching.

Because the ZnO layer is thin (i.e., 800 nm), they become fragile during KOH etching, and the resulting membranes tend to collapse on themselves. To avoid this problem, a 300 nm-thick layer of PMMA resin is deposited directly on the quartz surface (front face of the sample) using a spinner (3000 rpm, 30 sec) and then a hard bake at 140°C for 1 min before KOH etching. The resin maintains the quartz structure and is resistant to the etching solution for a given amount of time. The etching speed of silicon in KOH 45% at 80°C is 1 µm·min-1. The result is a transparent film of quartz and PMMA. The resin can be further removed using an $O_2$ plasma.

*Vibrometry Measurements at air and liquid cell conditions*

The 3D vibration reconstructed images and spectra of the fabricated quartz/silicon MEMS were evaluated by LDV equipped with laser, photodetector, and frequency generator. The vibrometer (MSA-600, Polytech®) allowed for characterizing in-plane and out-of-plane motion through non-contact silicon encapsulation without the need to prepare or decapsulate the device up to 25 MHz. Then, the dynamic displacement of the resonating MEMS sensor is reconstituted by the PSV software. This equipment has an internal frequency generator utilized to actuate the inverse piezoelectricity of quartz membranes.

*Quartz/silicon membrane MEMS manufacturing*

A 2-inch (100)Si wafer coated with the thin piezoelectric epitaxial quartz film. The wafer was opened using a laser to create its moving part. The power and repetition rate varies according to the thickness of the wafer and the desired depth of the moving part with different morphologies and sizes. The 2-inch silicon wafer is then placed in support from AMMT to protect the thin film of quartz before being immersed in a TMAH bath heated to 80°C by a water bath. A protective layer is applied by sputtering (PECVD) to the wafer surface, but on the other side, it is from the part covered with quartz. This protective layer is composed of three successive depositions: silicon nitride (400nm), silicon dioxide (400 nm), and again silicon nitride (400 nm). Then, the protective layer is etched until the silicon bulk is revealed ether by ICP-RIE with a $CHF_3/O_2$ plasma or with the laser etching system. If the ICP-RIE is chosen for the etching of the protective layer, a lithography step before the RIE etching is mandatory to have the desired membrane design. Then, a laser beam of 3.75 μm in diameter and 5 W in power is used to etch the membrane by uploading a designed membrane file in a laser machine. Due to this flexible and easy etching method, we can etch any shape of membrane we want in a few minutes. Before the TMAH (CAS sigma Aldricht number 75-59-2) etching step the native oxide layer that is formed on the silicon surface is removed by using Hydrofluoric acid (BOE 10% sigma Aldricht, CAS number 76664-39-3).

**Acknowledgments**

This project received funding from the European Research Council (ERC) under the European Union's Horizon 2020 research and innovation program (project SENSiSOFT No.803004. We also acknowledge financial support from Projects No. PID2020-118479RBI00 and PID2023-152225NB-I00. The authors thank Frederic Pichot, for the expertise and advice during the membranes lithographic processes at the CTM platform. Electron microscopy observations at ORNL were supported by the Office of Science, Materials Sciences and Engineering Division of the U.S. Department of Energy. The FEGSEM instrumentation was facilitated by the Institut des Matériaux de Paris Centre (IMPC FR2482) and was funded by Sorbonne Université, CNRS and by the C'Nano projects of the Région Ile-de-France. We thank David Montero for performing the FEGSEM images. Microscopy work was conducted at the Center for Nanophase Materials Sciences, which is a DOE Office of Science User Facility.

**Author contributions**

Conceptualization: A.C.-G, methodology: D.S.-F, A.R, R.D, L.G, S.B, S.D, N.C, F.P, R.G-B, N.G, G.A, J.G, C.M, S.P., and A.C.-G supervision: C.G, G.A., and A.C.-G, funding acquisition: A.C.-G, and C.G. writing: A.C.-G, with inputs from authors.

REFERENCES

(1) Ding, C.; Jia, H.; Sun, Q.; Yao, Z.; Yang, H.; Liu, W.; Pang, X.; Li, S.; Liu, C.; Minari, T.; Chen, J.; Liu, X.; Song, Y. Wafer-Scale Single Crystals: Crystal Growth Mechanisms, Fabrication Methods, and Functional Applications. *J. Mater. Chem. C* **2021**, *9* (25), 7829–7851. https://doi.org/10.1039/D1TC01101D.

(2) Kang, K.; Xie, S.; Huang, L.; Han, Y.; Huang, P. Y.; Mak, K. F.; Kim, C.-J.; Muller, D.; Park, J. High-Mobility Three-Atom-Thick Semiconducting Films with Wafer-Scale Homogeneity. *Nature* **2015**, *520* (7549), 656–660. https://doi.org/10.1038/nature14417.

(3) Vila-Fungueiriño, J. M.; Bachelet, R.; Saint-Girons, G.; Gendry, M.; Gich, M.; Gazquez, J.; Ferain, E.; Rivadulla, F.; Rodriguez-Carvajal, J.; Mestres, N.; Carretero-Genevrier, A. Integration of Functional Complex Oxide Nanomaterials on Silicon. *Frontiers in Physics* **2015**, *3*. https://doi.org/10.3389/fphy.2015.00038.

(4) France, R. M.; Geisz, J. F.; García, I.; Steiner, M. A.; McMahon, W. E.; Friedman, D. J.; Moriarty, T. E.; Osterwald, C.; Ward, J. S.; Duda, A.; Young, M.; Olavarria, W. J. Design Flexibility of Ultrahigh Efficiency Four-Junction Inverted Metamorphic Solar Cells. *IEEE Journal of Photovoltaics* **2016**, *6* (2), 578–583. https://doi.org/10.1109/JPHOTOV.2015.2505182.

(5) Tian, P.; McKendry, J. J. D.; Gong, Z.; Zhang, S.; Watson, S.; Zhu, D.; Watson, I. M.; Gu, E.; Kelly, A. E.; Humphreys, C. J.; Dawson, M. D. Characteristics and Applications of Micro-Pixelated GaN-Based Light Emitting Diodes on Si Substrates. *Journal of Applied Physics* **2014**, *115* (3), 033112. https://doi.org/10.1063/1.4862298.

(6) Cerba, T.; Martin, M.; Moeyaert, J.; David, S.; Rouviere, J. L.; Cerutti, L.; Alcotte, R.; Rodriguez, J. B.; Bawedin, M.; Boutry, H.; Bassani, F.; Bogumilowicz, Y.; Gergaud, P.; Tournié, E.; Baron, T. Anti Phase Boundary Free GaSb Layer Grown on 300mm (001)-Si Substrate by Metal Organic Chemical Vapor Deposition. *Thin Solid Films* **2018**, *645*, 5–9. https://doi.org/10.1016/j.tsf.2017.10.024.

(7) Tournié, E.; Cerutti, L.; Rodriguez, J.-B.; Liu, H.; Wu, J.; Chen, S. Metamorphic III–V Semiconductor Lasers Grown on Silicon. *MRS Bulletin* **2016**, *41* (3), 218–223. https://doi.org/10.1557/mrs.2016.24.

(8) Fathpour, S. Emerging Heterogeneous Integrated Photonic Platforms on Silicon. *Nanophotonics* **2015**, *4* (2), 143–164. https://doi.org/doi:10.1515/nanoph-2014-0024.

(9) Isarakorn, D.; Sambri, A.; Janphuang, P.; Briand, D.; Gariglio, S.; Triscone, J.-M.; Guy, F.; Reiner, J. W.; Ahn, C. H.; de Rooij, N. F. Epitaxial piezoelectric MEMS on silicon. *J. Micromech. Microeng.* **2010**, *20* (5), 055008. https://doi.org/10.1088/0960-1317/20/5/055008.

(10) Carretero-Genevrier, A.; Oro-Sole, J.; Gazquez, J.; Magen, C.; Miranda, L.; Puig, T.; Obradors, X.; Ferain, E.; Sanchez, C.; Rodriguez-Carvajal, J.; Mestres, N. Direct Monolithic

Integration of Vertical Single Crystalline Octahedral Molecular Sieve Nanowires on Silicon. *Chemistry of Materials* **2014**, *26* (2), 1019–1028. https://doi.org/10.1021/cm403064u.

(11) Gomez, A.; Vila-Fungueiriño, J. M.; Jolly, C.; Garcia-Bermejo, R.; Oró-Solé, J.; Ferain, E.; Mestres, N.; Magén, C.; Gazquez, J.; Rodriguez-Carvajal, J.; Carretero-Genevrier, A. Crystal Engineering and Ferroelectricity at the Nanoscale in Epitaxial 1D Manganese Oxide on Silicon. *Nanoscale* **2021**, *13* (21), 9615–9625. https://doi.org/10.1039/D1NR00565K.

(12) McKee, R. A.; Walker, F. J.; Chisholm, M. F. Crystalline Oxides on Silicon: The First Five Monolayers. *Physical Review Letters* **1998**, *81* (14), 3014–3017. https://doi.org/10.1103/PhysRevLett.81.3014.

(13) Gómez, A.; Vila-Fungueiriño, J. M.; Moalla, R.; Saint-Girons, G.; Gázquez, J.; Varela, M.; Bachelet, R.; Gich, M.; Rivadulla, F.; Carretero-Genevrier, A. Electric and Mechanical Switching of Ferroelectric and Resistive States in Semiconducting BaTiO3–δ Films on Silicon. *Small* n/a-n/a. https://doi.org/10.1002/smll.201701614.

(14) Saint-Girons, G.; Bachelet, R.; Moalla, R.; Meunier, B.; Louahadj, L.; Canut, B.; Carretero-Genevrier, A.; Gazquez, J.; Regreny, P.; Botella, C.; Penuelas, J.; Silly, M. G.; Sirotti, F.; Grenet, G. Epitaxy of SrTiO3 on Silicon: The Knitting Machine Strategy. *Chemistry of Materials* **2016**. https://doi.org/10.1021/acs.chemmater.6b01260.

(15) Lapano, J.; Brahlek, M.; Zhang, L.; Roth, J.; Pogrebnyakov, A.; Engel-Herbert, R. Scaling Growth Rates for Perovskite Oxide Virtual Substrates on Silicon. *Nature Communications* **2019**, *10* (1), 2464. https://doi.org/10.1038/s41467-019-10273-2.

(16) Gu, X.; Lubyshev, D.; Batzel, J.; Fastenau, J. M.; Liu, W. K.; Pelzel, R.; Magana, J. F.; Ma, Q.; Rao, V. R. Growth, Characterization, and Uniformity Analysis of 200 Mm Wafer-Scale SrTiO3/Si. *Journal of Vacuum Science & Technology B* **2010**, *28* (3), C3A12-C3A16. https://doi.org/10.1116/1.3292509.

(17) Vila-Fungueiriño, J. M.; Gázquez, J.; Magén, C.; Saint-Girons, G.; Bachelet, R.; Carretero-Genevrier, A. Epitaxial La0.7Sr0.3MnO3 Thin Films on Silicon with Excellent Magnetic and Electric Properties by Combining Physical and Chemical Methods. *Science and Technology of Advanced Materials* **2018**, *19* (1), 702–710. https://doi.org/10.1080/14686996.2018.1520590.

(18) Janotti, A.; Walle, C. G. V. de. Fundamentals of Zinc Oxide as a Semiconductor. *Reports on Progress in Physics* **2009**, *72* (12), 126501. https://doi.org/10.1088/0034-4885/72/12/126501.

(19) Tsukazaki, A.; Ohtomo, A.; Kita, T.; Ohno, Y.; Ohno, H.; Kawasaki, M. Quantum Hall Effect in Polar Oxide Heterostructures. *Science* **2007**, *315* (5817), 1388. https://doi.org/10.1126/science.1137430.

(20) Lu, X.; Wang, G.; Xie, S.; Shi, J.; Li, W.; Tong, Y.; Li, Y. Efficient Photocatalytic Hydrogen Evolution over Hydrogenated ZnO Nanorod Arrays. *Chem. Commun.* **2012**, *48* (62), 7717–7719. https://doi.org/10.1039/C2CC31773G.

(21) Tian, C.; Zhang, Q.; Wu, A.; Jiang, M.; Liang, Z.; Jiang, B.; Fu, H. Cost-Effective Large-Scale Synthesis of ZnO Photocatalyst with Excellent Performance for Dye Photodegradation. *Chem. Commun.* **2012**, *48* (23), 2858–2860. https://doi.org/10.1039/C2CC16434E.

(22) Nandi, S.; Kumar, S.; Misra, A. Zinc Oxide Heterostructures: Advances in Devices from Self-Powered Photodetectors to Self-Charging Supercapacitors. *Mater. Adv.* **2021**, *2* (21), 6768–6799. https://doi.org/10.1039/D1MA00670C.

(23) Wang, Z. L.; Song, J. Piezoelectric Nanogenerators Based on Zinc Oxide Nanowire Arrays. *Science* **2006**, *312* (5771), 242. https://doi.org/10.1126/science.1124005.

(24) Zhao, J.; Hu, L.; Wang, Z.; Wang, Z.; Zhang, H.; Zhao, Y.; Liang, X. Epitaxial Growth of ZnO Thin Films on Si Substrates by PLD Technique. *Journal of Crystal Growth* **2005**, *280* (3), 455–461. https://doi.org/10.1016/j.jcrysgro.2005.03.071.

(25) Nahhas, A.; Kim, H. K.; Blachere, J. Epitaxial Growth of ZnO Films on Si Substrates Using an Epitaxial GaN Buffer. *Applied Physics Letters* **2001**, *78* (11), 1511–1513. https://doi.org/10.1063/1.1355296.

(26) Adrien Carretero-Genevrier; David Sanchez-Fuentes, Lorenzo Garcia, Ricardo Garcia, Samir Bousri, Javier Moral-Vico. Piezoelectric Epitaxially Grown Pseudo Substrate, Use and Process for Preparing Such a Pseudosubstrate, PCT/FR2022/051466. WO2023002139A1. PCT/FR2022/051466.

(27) Bindini, E.; Naudin, G.; Faustini, M.; Grosso, D.; Boissière, C. The Critical Role of the Atmosphere in Dip-Coating Process. *Journal of Physical Chemistry C* **2017**, *121*, 14572–14580.

(28) Brinker, C. J.; Lu, Y. F.; Sellinger, A.; Fan, H. Y. Evaporation-Induced Self-Assembly: Nanostructures Made Easy. *Advanced Materials* **1999**, *11* (7), 579-+. https://doi.org/10.1002/(sici)1521-4095(199905)11:7<579::aid-adma579>3.0.co;2-r.

(29) Carretero-Genevrier, A.; Gich, M.; Picas, L.; Gazquez, J.; Drisko, G. L.; Boissiere, C.; Grosso, D.; Rodriguez-Carvajal, J.; Sanchez, C. Soft-Chemistry-Based Routes to Epitaxial Alpha-Quartz Thin Films with Tunable Textures. *Science* **2013**, *340* (6134), 827–831. https://doi.org/10.1126/science.1232968.

(30) Bitenc, M.; Podbršček, P.; Dubček, P.; Bernstorff, S.; Dražić, G.; Orel, B.; Orel, Z. C. The Growth Mechanism of Zinc Oxide and Hydrozincite: A Study Using Electron Microscopies and in Situ SAXS. *CrystEngComm* **2012**, *14* (9), 3080–3088. https://doi.org/10.1039/C2CE06134A.

(31) Stählin, W.; Oswald, H. R. The Crystal Structure of Zinc Hydroxide Nitrate, Zn\sb 5(OH)\sb 8(NO\sb 3)\sb 2.2H\sb 2O. *Acta Crystallographica Section B* **1970**, *26* (6), 860–863. https://doi.org/10.1107/S0567740870003230.

(32) Emanetoglu, N. W.; Liang, S.; Gorla, C.; Lu, Y.; Jen, S.; Subramanian, R. Epitaxial Growth and Characterization of High Quality ZnO Films for Surface Acoustic Wave Applications. In *1997 IEEE Ultrasonics Symposium Proceedings. An International Symposium (Cat. No.97CH36118)*; 1997; Vol. 1, pp 195–200 vol.1. https://doi.org/10.1109/ULTSYM.1997.663009.

(33) Zhu, G.; Yang, R.; Wang, S.; Wang, Z. L. Flexible High-Output Nanogenerator Based on Lateral ZnO Nanowire Array. *Nano Lett.* **2010**, *10* (8), 3151–3155. https://doi.org/10.1021/nl101973h.

(34) Sun, Y.; Chen, L.; Bao, Y.; Zhang, Y.; Wang, J.; Fu, M.; Wu, J.; Ye, D. The Applications of Morphology Controlled ZnO in Catalysis. *Catalysts* **2016**, *6* (12). https://doi.org/10.3390/catal6120188.

(35) Luo, J.-T.; Quan, A.-J.; Zheng, Z.-H.; Liang, G.-X.; Li, F.; Zhong, A.-H.; Ma, H.-L.; Zhang, X.-H.; Fan, P. Study on the Growth of Al-Doped ZnO Thin Films with (1$\bar{1}$0) and (0002) Preferential Orientations and Their Thermoelectric Characteristics. *RSC Adv.* **2018**, *8* (11), 6063–6068. https://doi.org/10.1039/C7RA12485F.

(36) Wu, Y.; Zhang, L.; Xie, G.; Ni, J.; Chen, Y. Structural and Electrical Properties of (110) ZnO Epitaxial Thin Films on (001) SrTiO3 Substrates. *Solid State Communications* **2008**, *148* (5), 247–250. https://doi.org/10.1016/j.ssc.2008.08.009.

(37) Ho, Y.-T.; Wang, W.-L.; Peng, C.-Y.; Liang, M.-H.; Tian, J.-S.; Lin, C.-W.; Chang, L. Growth of Nonpolar (11$\bar{2}$0) ZnO Films on LaAlO3 (001) Substrates. *Appl. Phys. Lett.* **2008**, *93* (12), 121911. https://doi.org/10.1063/1.2988167.

(38) Wang, Y.; Zhang, S.; Wasa, K.; Shui, X. Deposition of ZnO Films with C-Axes Lying in R-Sapphire Substrate Planes. *The Journal of the Acoustical Society of America* **2012**, *131* (4), 3466–3466. https://doi.org/10.1121/1.4709066.

(39) Xiao, Y.; Tian, Y.; Sun, S.; Chen, C.; Wang, B. Growth Modulation of Simultaneous Epitaxy of ZnO Obliquely Aligned Nanowire Arrays and Film on R-Plane Sapphire Substrate. *Nano Research* **2018**, *11* (7), 3864–3876. https://doi.org/10.1007/s12274-017-1960-1.

(40) Gemmi, M.; Mugnaioli, E.; Gorelik, T. E.; Kolb, U.; Palatinus, L.; Boullay, P.; Hovmöller, S.; Abrahams, J. P. 3D Electron Diffraction: The Nanocrystallography Revolution. *ACS Cent. Sci.* **2019**, *5* (8), 1315–1329. https://doi.org/10.1021/acscentsci.9b00394.

(41) Plana-Ruiz, S.; Krysiak, Y.; Portillo, J.; Alig, E.; Estradé, S.; Peiró, F.; Kolb, U. Fast-ADT: A Fast and Automated Electron Diffraction Tomography Setup for Structure Determination and Refinement. *Ultramicroscopy* **2020**, *211*, 112951. https://doi.org/10.1016/j.ultramic.2020.112951.

(42) Kelley, K. P.; Yilmaz, D. E.; Collins, L.; Sharma, Y.; Lee, H. N.; Akbarian, D.; van Duin, A. C. T.; Ganesh, P.; Vasudevan, R. K. Thickness and Strain Dependence of Piezoelectric Coefficient in \mathrmBaTiO_3 Thin Films. *Phys. Rev. Mater.* **2020**, *4* (2), 024407. https://doi.org/10.1103/PhysRevMaterials.4.024407.

(43) Mustaffa, M. A.; Arith, F.; Noorasid, N. S.; Zin, M. S. I. M.; Leong, K. S.; Ali, F. A.; Mustafa, A. N. M.; Ismail, M. M. Towards a Highly Efficient ZnO Based Nanogenerator. *Micromachines* **2022**, *13* (12). https://doi.org/10.3390/mi13122200.

(44) Yang, R.; Qin, Y.; Dai, L.; Wang, Z. L. Power Generation with Laterally Packaged Piezoelectric Fine Wires. *Nature Nanotechnology* **2009**, *4* (1), 34–39. https://doi.org/10.1038/nnano.2008.314.

(45) Plana-Ruiz, S.; Portillo, J.; Estradé, S.; Peiró, F.; Kolb, U.; Nicolopoulos, S. Quasi-Parallel Precession Diffraction: Alignment Method for Scanning Transmission Electron Microscopes. *Ultramicroscopy* **2018**, *193*, 39–51. https://doi.org/10.1016/j.ultramic.2018.06.005.

(46) Kolb, U.; Gorelik, T.; Kübel, C.; Otten, M. T.; Hubert, D. Towards Automated Diffraction Tomography: Part I--Data Acquisition. *Ultramicroscopy* **2007**, *107* (6–7), 507–513. https://doi.org/10.1016/j.ultramic.2006.10.007.

(47) Palatinus, L.; Brázda, P.; Jelínek, M.; Hrdá, J.; Steciuk, G.; Klementová, M. Specifics of the Data Processing of Precession Electron Diffraction Tomography Data and Their Implementation in the Program PETS2.0. *Acta Crystallographica Section B* **2019**, *75* (4), 512–522. https://doi.org/10.1107/S2052520619007534.

(48) Petříček, V.; Palatinus, L.; Plášil, J.; Dušek, M. Jana2020 – a New Version of the Crystallographic Computing System Jana. *Zeitschrift für Kristallographie - Crystalline Materials* **2023**, *238* (7–8), 271–282. https://doi.org/doi:10.1515/zkri-2023-0005.

(49) Palatinus, L.; Petříček, V.; Corrêa, C. A. Structure Refinement Using Precession Electron Diffraction Tomography and Dynamical Diffraction: Theory and Implementation. *Acta Crystallogr A Found Adv* **2015**, *71* (Pt 2), 235–244. https://doi.org/10.1107/S2053273315001266.

(50) Ophus, C. Four-Dimensional Scanning Transmission Electron Microscopy (4D-STEM): From Scanning Nanodiffraction to Ptychography and Beyond. *Microscopy and Microanalysis* **2019**, *25* (3), 563–582. https://doi.org/10.1017/S1431927619000497.

(51) Bui, Q. C.; Ardila, G.; Sarigiannidou, E.; Roussel, H.; Jiménez, C.; Chaix-Pluchery, O.; Guerfi, Y.; Bassani, F.; Donatini, F.; Mescot, X.; Salem, B.; Consonni, V. Morphology Transition of ZnO from Thin Film to Nanowires on Silicon and Its Correlated Enhanced Zinc Polarity Uniformity and Piezoelectric Responses. *ACS Appl. Mater. Interfaces* **2020**, *12* (26), 29583–29593. https://doi.org/10.1021/acsami.0c04112.

(52) Savitzky, B. H.; Zeltmann, S. E.; Hughes, L. A.; Brown, H. G.; Zhao, S.; Pelz, P. M.; Pekin, T. C.; Barnard, E. S.; Donohue, J.; Rangel DaCosta, L.; Kennedy, E.; Xie, Y.; Janish, M. T.; Schneider, M. M.; Herring, P.; Gopal, C.; Anapolsky, A.; Dhall, R.; Bustillo, K. C.; Ercius, P.; Scott, M. C.; Ciston, J.; Minor, A. M.; Ophus, C. py4DSTEM: A Software Package for Four-Dimensional Scanning Transmission Electron Microscopy Data Analysis. *Microsc Microanal* **2021**, *27* (4), 712–743. https://doi.org/10.1017/S1431927621000477.